\documentclass[aps,prb,twocolumn,showpacs,groupedaddress]{revtex4}

\usepackage{graphicx}
\usepackage{dcolumn}

\begin{document}


\title{
Quantum effects in 
small-capacitance single Josephson junctions
}


\author{Michio Watanabe}
\email[]{michio@postman.riken.go.jp}
\homepage[]{http://www.riken.go.jp/lab-www/semiconductors/michio/}
\affiliation{
Semiconductors Laboratory, RIKEN (The Institute 
of Physical and Chemical Research), 2-1 Hirosawa, 
Wako-shi, Saitama 351-0198, Japan
}
\author{David B. Haviland}
\email[]{haviland@nanophys.kth.se}
\homepage[]{http://www.nanophys.kth.se/}
\affiliation{
Nanostructure Physics, The Royal Institute of 
Technology (KTH), SCFAB, Roslagstullsbacken 21, 
106 91 Stockholm, Sweden
}


\date{2 August 2002}

\begin{abstract}
We have measured the current-voltage ($I$-$V$) characteristics 
of small-capacitance single Josephson junctions 
at low temperatures ($T=0.02-0.6$~K),  
where the strength of the coupling 
between the single junction and the electromagnetic environment 
was controlled with one-dimensional arrays of dc 
superconducting quantum interference devices (SQUIDs).  
The single-junction $I$-$V$ curve is sensitive to the impedance 
of the environment, which can be tuned {\itshape in situ}.
We have observed Coulomb blockade of Cooper-pair tunneling 
and even a region of negative differential resistance, 
when the zero-bias resistance $R_0'$ of the SQUID arrays 
is much higher than the quantum resistance 
$R_K\equiv h/e^2\approx26$~k$\Omega$.  
The negative differential resistance is evidence 
of the coherent single-Cooper-pair tunneling 
within the theory of current-biased single 
Josephson junctions.  Based on this theory, we have calculated 
the $I$-$V$ curves numerically in order to compare them with the 
experimental ones at $R_0'\gg R_K$.  The numerical calculation 
agrees with the experiments qualitatively.  
We also discuss the $R_0'$ dependence 
of the single-Josephson-junction $I$-$V$ curve 
in terms of the superconductor-insulator 
transition driven by changing the coupling to the environment.  
\end{abstract}


\maketitle

\section{Introduction}
\label{sec:intro}
Small-capacitance superconducting tunnel junctions provide 
an ideal system for studying 
the interplay between quantum mechanically conjugate variables:   
the Josephson-phase difference across the junction 
and the charge on the junction electrode.  
Their behavior is influenced by dissipation, 
such as discrete tunneling of quasiparticles, 
coupling to an electromagnetic environment, etc.
The simplest and the most fundamental example is the single junction.  
Current-voltage ($I$-$V$) characteristics of single junctions have 
been the subject of extensive theoretical investigations.\cite{Ave91,Sch90}  
A small capacitance $C$ gives a large charging energy $e^2/2C$, and thus 
the theory predicts that at low enough temperatures,  
a {\itshape superconducting} small-capacitance tunnel junction behaves 
like an {\itshape insulator} so long as the junction is isolated 
from the electromagnetic environment.  
Experimentally, however, the observation of this charging effect 
(Coulomb blockade) in a single junction is not straightforward, 
because the measurement leads attached to the single junction 
should have a high enough impedance  
to isolate the junction.\cite{Ave91}  
For this reason, thin-film resistors were employed for the leads, 
and Coulomb blockade was successfully observed in single 
Josephson junctions.\cite{Hav91}
Tunnel-junction arrays were also tried for the leads, and an increase 
of differential resistance around $V=0$ was reported.\cite{Gee90,Shi97}   

We use one-dimensional (1D) arrays of dc superconducting quantum 
interference devices (SQUIDs) for the leads.\cite{Wat01PRL}  
The advantage of this SQUID configuration is that the effective impedance 
of an array can be varied {\it in situ} by applying an external magnetic 
field perpendicular to the SQUID loop.  Thus, the zero-bias resistance 
of the SQUID arrays at low temperatures can be controlled  
over several orders of magnitude.  
This phenomenon has been extensively studied in terms of 
the superconductor-insulator (SI) 
transition.\cite{Cho98,Hav00,Hav01,Wat02}  
The single junction in our samples, on the other hand, 
does not have a SQUID configuration, 
and therefore its parameters are practically independent 
of the external magnetic field.  This enables us to study 
the {\it same} single junction in {\itshape different} environments.  
We show that the $I$-$V$ curve of the single junction is indeed sensitive 
to the state of the environment.  Furthermore, we can induce 
a transition to a Coulomb blockade of the single junction 
by increasing the zero-bias resistance $R_0'$ of the SQUID arrays.  
When $R_0'$ is much higher 
than the quantum resistances: $R_K\equiv h/e^2\approx26$~k$\Omega$ 
for quasiparticles and $R_Q\equiv h/(2e)^2\approx6.5$~k$\Omega$ 
for Cooper pairs, single-junction $I$-$V$ curve has a region 
of negative differential resistance.  The negative differential 
resistance is a result of coherent single-Cooper-pair tunneling 
from the viewpoint of the theory for current-biased single 
Josephson junctions whose electromagnetic environment has 
a sufficiently high impedance.\cite{Ave91,Sch90} 

In the context of the theory, we calculate the $I$-$V$ 
curves numerically by following Ref.~\onlinecite{Gei88}, 
and compare with the experimental ones at $R_0'\gg R_K$.   
The numerical results are in qualitative agreement 
with the experiments,\cite{Wat01SUST} however, 
some quantitative discrepancies suggest that 
even at $R_0'\gg R_K$, the single-junction $I$-$V$ curve 
would be influenced by the environment.  
According to the perturbation theory\cite{Fal91,Ing92,Ing99} 
that deals with the effect of a linear-impedance environment 
$Z(\omega)$ on single Josephson junctions, 
$\mbox{Re}[Z(\omega)]$, or the dissipation, plays a key role.  
An insulating behavior characterized by the Coulomb blockade 
of Cooper-pair tunneling is expected for the single junction 
when $\mbox{Re}[Z(\omega)]\gg R_Q$, i.e., the dissipation is 
sufficiently small, because the large $\mbox{Re}[Z(\omega)]$ 
suppresses the charge fluctuation.   
The dissipative dynamics of single Josephson junctions has also 
been discussed by studying resistively shunted junctions 
both theoretically\cite{Sch90,Kat00,Chu02} and 
experimentally.\cite{Yag97,Pen01} 
As the dissipation is increased (the shunt resistance is decreased), 
the tunneling of the Josephson phase is suppressed, and as a result 
the Josephson junction exhibits a transition from an insulator to 
a superconductor.  The theoretical phase diagram\cite{Sch90} 
for this SI transition was supported by the 
experiments\cite{Yag97,Pen01} qualitatively.  
The superconducting behavior has been observed experimentally 
in single Josephson junctions not only by shunting with a resistor 
but also by biasing with a low-impedance circuit,\cite{Ste01}  
i.e., by achieving a low-impedance electromagnetic environment.  
Thus, we expect that there should be a SI transition 
driven by the electromagnetic environment.  
We discuss the environment-driven SI transition in our system, 
where the environment is tunable with the SQUID arrays.  

\section{Theory}
\begin{figure}
\includegraphics[width=0.95\columnwidth,clip]{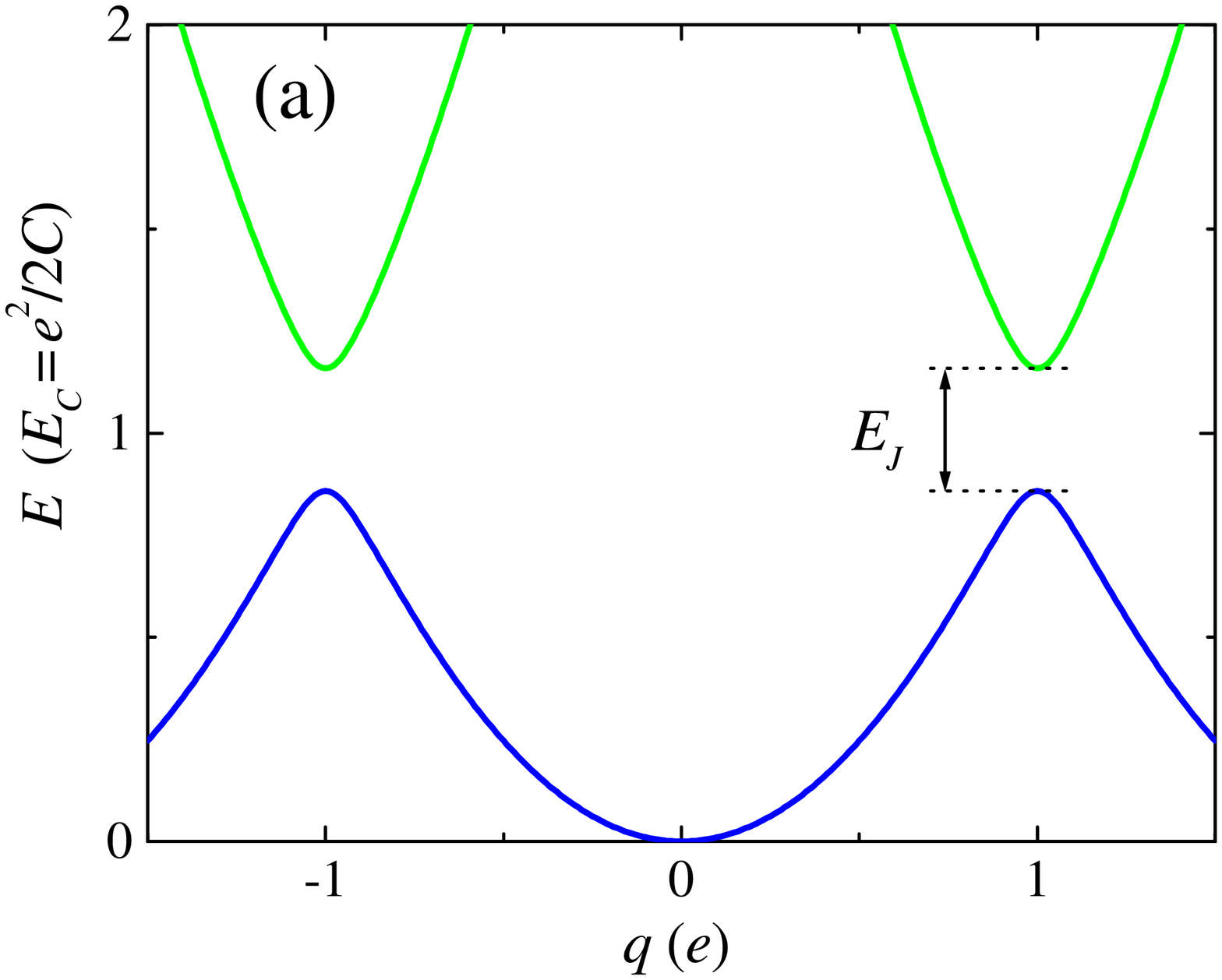}

\vspace{\baselineskip}

\includegraphics[width=0.95\columnwidth,clip]{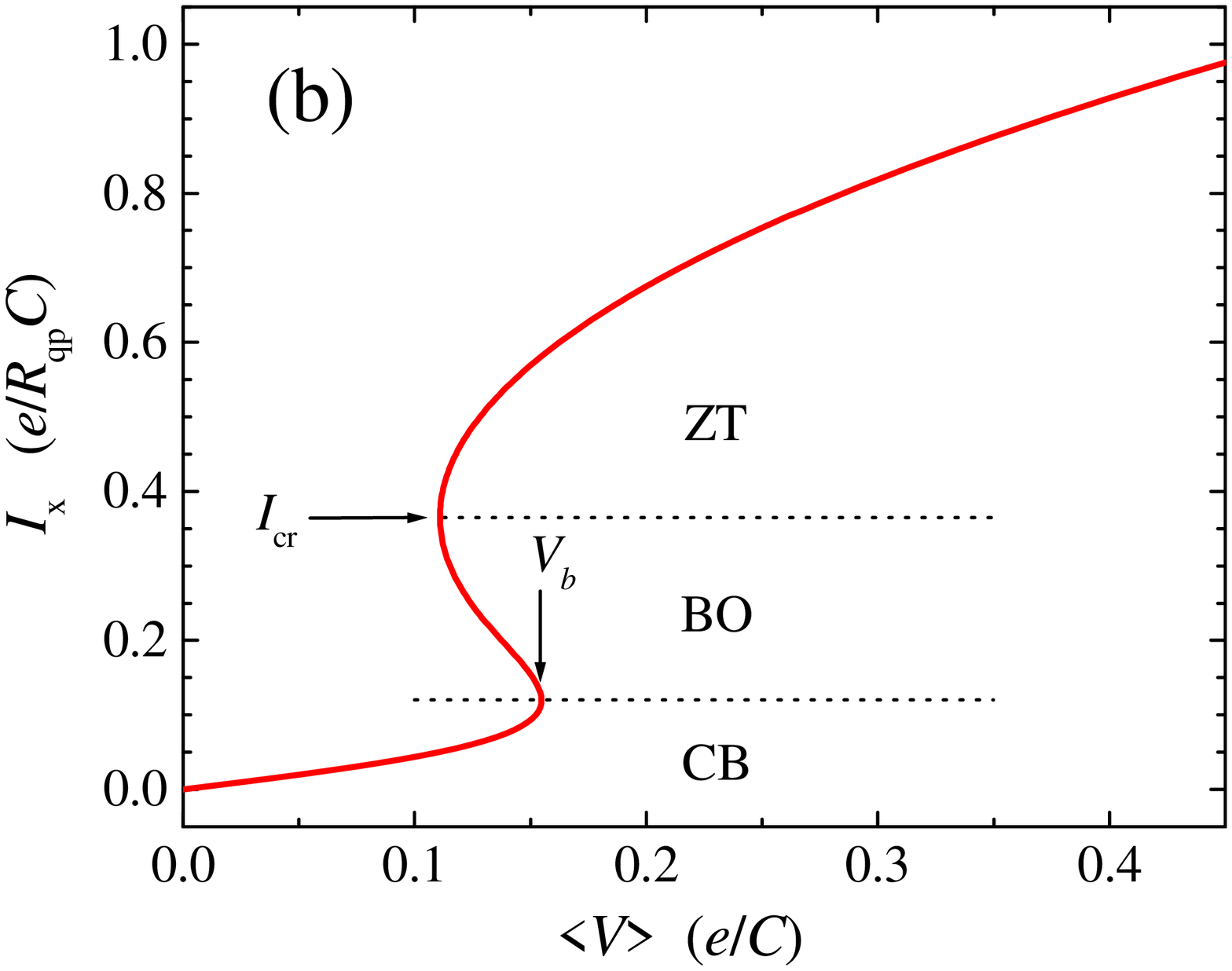}
\caption{\label{fig:calc}
(a) Energy diagram (energy eigenvalue in units of $E_C$ vs. quasicharge) 
and (b) theoretical current-voltage characteristics 
for a single Josephson junction with $E_J/E_C=0.3$ and $R_{\rm qp}
=10^2\,(h/\pi^2e^2)$ at $k_BT/E_C=0.3$, where $E_J$ is the Josephson 
energy, $E_C\equiv e^2/2C$ is the charging energy, $R_{\rm qp}$ is the 
quasiparticle resistance, and $k_BT$ is the thermal energy.    
(The energy diagram depends only on $E_J/E_C$.)  
}
\end{figure}

The Hamiltonian of a single Josephson junction in an environment 
with sufficiently high impedance is written as 
\begin{equation}
\label{eq:band1}
H = \frac{Q^2}{2C} - E_J\cos\phi\,, 
\end{equation}
where $Q$ is the charge on the junction electrode, 
$C$ is the capacitance of the junction, 
$E_J$ is the Josephson energy, and $\phi$ is 
the Josephson-phase difference across the junction.  
The charge $Q$ and $\hbar\phi/2e$ are quantum-mechanically 
conjugate valuables, and a set of the eigenfunctions are 
Bloch waves of the form 
\begin{equation}
\psi(\phi)=u(\phi)\exp(i\phi q/2e)\,,    
\end{equation}
where $q$ is called quasicharge 
and $u(\phi)$ is a periodic function, 
\begin{equation}
u(\phi+2\pi)=u(\phi)\,.
\end{equation}
The energy eigenvalue $E$ plotted 
as a function of $q$ has a band structure, 
and in all the allowed bands, it is $2e$ periodic. 
An example of the energy diagram for $E_J/E_C=0.3$, 
where $E_C\equiv e^2/2C$ is the charging energy, 
is shown in Fig.~\ref{fig:calc}a.    
Under constant current bias $I_{\rm x}$, 
in the absence of quasiparticle or Cooper-pair tunneling, 
$q$ increases uniformly in time according to 
\begin{equation}
\label{eq:IV1}
\frac{dq}{dt}=I_{\rm x}\,,
\end{equation}
so that the state of the system advances toward higher $q$ 
within a given band as time goes on.  
The average voltage is given by 
\begin{equation}
\label{eq:IV6}
\langle V\rangle=\sum_{i_b,q}P(i_b,q)\,\frac{\,dE(i_b,q)\,}{dq}\,,
\end{equation}
where $i_b$ is the band index and $P(i_b,q)$ is the probability 
that the system is in the state $(i_b,q)$.  
The probability $P(i_b,q)$ can be calculated by solving  
a set of coupled differential equations of the form 
\begin{equation}
\label{eq:IV8}
\frac{\,dP(i_b,q)\,}{dt}
=\sum_{i_b',q'}A(i_b,q,i_b',q')\,P(i_b',q')=0~,
\end{equation}
where the matrix element $A(i_b,q,i_b',q')$ describes 
the rate of transition between the states $(i_b,q)$ 
and $(i_b',q')$.  The dominant process for 
$A(i_b,q,i_b',q')$ depends on the magnitude of $I_{\rm x}$.  

An example of the theoretical $I$-$V$ curve is shown in 
Fig.~\ref{fig:calc}b.  
For sufficiently small $I_{\rm x}$ (region~CB), the dominant process 
is stochastic quasiparticle tunneling, where $q$ changes by $e$.  
This tunneling always occurs along the energy parabola $q^2/2C$, 
such that $i_b$ changes by 0 or $+1$ if the initial state is 
in the lowest band ($i_b=1$) and by $\pm1$ 
for all the other initial states.\cite{Sch90}  
The rate for the quasiparticle tunneling is given by 
\begin{equation}
\label{eq:IV2}
\Gamma(\Delta E)=\frac{\Delta E/e^2R_{\rm qp}}
{\,\exp(\Delta E/k_BT)-1\,}~, 
\end{equation}
where $R_{\rm qp}$ is the quasiparticle resistance and 
$\Delta E$ is the difference in energy between the initial 
$(i_b,q)$ and final $(i_b',q')$ states, 
\begin{equation}
\label{eq:IV3}
\Delta E\equiv E(i_b',q')- E(i_b,q)~.  
\end{equation}
At sufficiently low temperatures, the tunneling with 
$\Delta E>0$ is extremely unfavorable, and the $I$-$V$ curve 
is highly resistive (Coulomb blockade, region~CB 
in Fig.~\ref{fig:calc}b).  
For larger $I_{\rm x}$ (region~BO in Fig.~\ref{fig:calc}b), 
the quasicharge is frequently driven 
to the boundary of the Brillouin zone, $q=e$, then taken to $-e$ 
as a Cooper pair tunnels (Bloch oscillation).  
This process decreases $\langle V\rangle$, 
and as a result the $I$-$V$ curve has a region of negative 
differential resistance, or ``back bending," in the low-current part.  
For still larger $I_{\rm x}$ (region~ZT in Fig.~\ref{fig:calc}b), 
Zener tunneling becomes important, and $\langle V\rangle$ 
increases again.  
In Zener tunneling, no quasicharge is transferred but the state of the 
system jumps from one band to another as it passes by the gap 
between the bands.  The probability of Zener tunneling from band $i_b$ 
to $i_b+1$ or vice versa is given by 
\begin{equation}
\label{eq:IV4}
P_Z=\exp\left[-\frac{\,\pi\,}{8}\frac{\,(\Delta E)^2\,}{\,i_bE_C\,}
\frac{e}{\,\hbar I_{\rm x}\,}\right]~.
\end{equation}
Following Ref.~\onlinecite{Gei88}
which takes into account the above tunneling processes (quasiparticle, 
Cooper pair, and Zener), we have calculated the $I$-$V$ curve 
numerically.\footnote{The numerical code was written by R. L. Kautz.}     
The parameters for the calculation are $E_J/E_C$, $k_BT/E_C$, and 
\begin{equation}
\label{eq:alpha}
\alpha\equiv \frac{h}{\,\pi^2e^2R_{\rm qp}\,}~.  
\end{equation}
The current and the voltage 
are in units of $e/R_{\rm qp}C$ and $e/C$, respectively.    

As we have seen in Fig.~\ref{fig:calc}b, a typical $I$-$V$ curve 
consists of three regions, so that it is characterized by the local 
voltage maximum, or blockade voltage $V_b$, and the local current 
minimum, or crossover current $I_{\rm cr}$.  
(Here, we have to mention that the back-bending feature is blurred out 
if $k_BT/E_C$ or $\alpha$ is increased considerably.)  
An analytical expression of $V_b$ and $I_{\rm cr}$ has been obtained 
theoretically for limiting cases.\cite{Sch90}  
The value of $V_b$ is a function of $E_J/E_C$, and given by   
\begin{equation}
V_b \approx 
\left\{
\begin{array}{ll} 
0.25\,e/C\;\; & \mbox{for $E_J/E_C \ll 1$} \\
\delta_0/e & \mbox{for $E_J/E_C \gg 1$,}
\end{array}
\right.
\end{equation}
as $T\to 0$, where 
\begin{equation}
\delta_0 = \frac{\,e^2\,}{C}\,8\left(\frac{1}{2\pi^2}\right)^{\!1/4}\!
\left(\frac{E_J}{E_C}\right)^{\!3/4}\exp\left[-\left(8\,\frac{E_J}{E_C}
\right)^{\!1/2}\,\right]
\end{equation}
is the half-width of the lowest energy-band.   
As for $I_{\rm cr}$, 
\begin{equation}
\label{eq:Icr}
I_{\rm cr}\sim\left(I_Z\,\,\frac{e}{\,R_{\rm qp}C\,}\right)^{1/2}
\end{equation} 
is expected for $\alpha\ll(E_J/E_C)^2\ll1$ and $T\rightarrow0$, where 
\begin{equation}
I_Z\equiv\frac{\,\pi\,}{8}\frac{eE_J^{\,\,2}}{\,\hbar E_C\,}
\end{equation} 
is the Zener breakdown current.  Note that $I_{\rm cr}$ is much smaller 
than $I_Z$.   
When we compare our experimental results with the theory, we need 
theoretical prediction for finite $k_BT/E_C$, and arbitrary 
$E_J/E_C$ and $\alpha$.  For this reason we have done the numerical 
calculation.  The measured $V_b$ and $I_{\rm cr}$ are compared 
with the calculation in Sec.~\ref{subsec:comp}.

\section{Experiment}
\subsection{Sample preparation and characterization}
\begin{figure}
\includegraphics[width=0.9\columnwidth,clip]{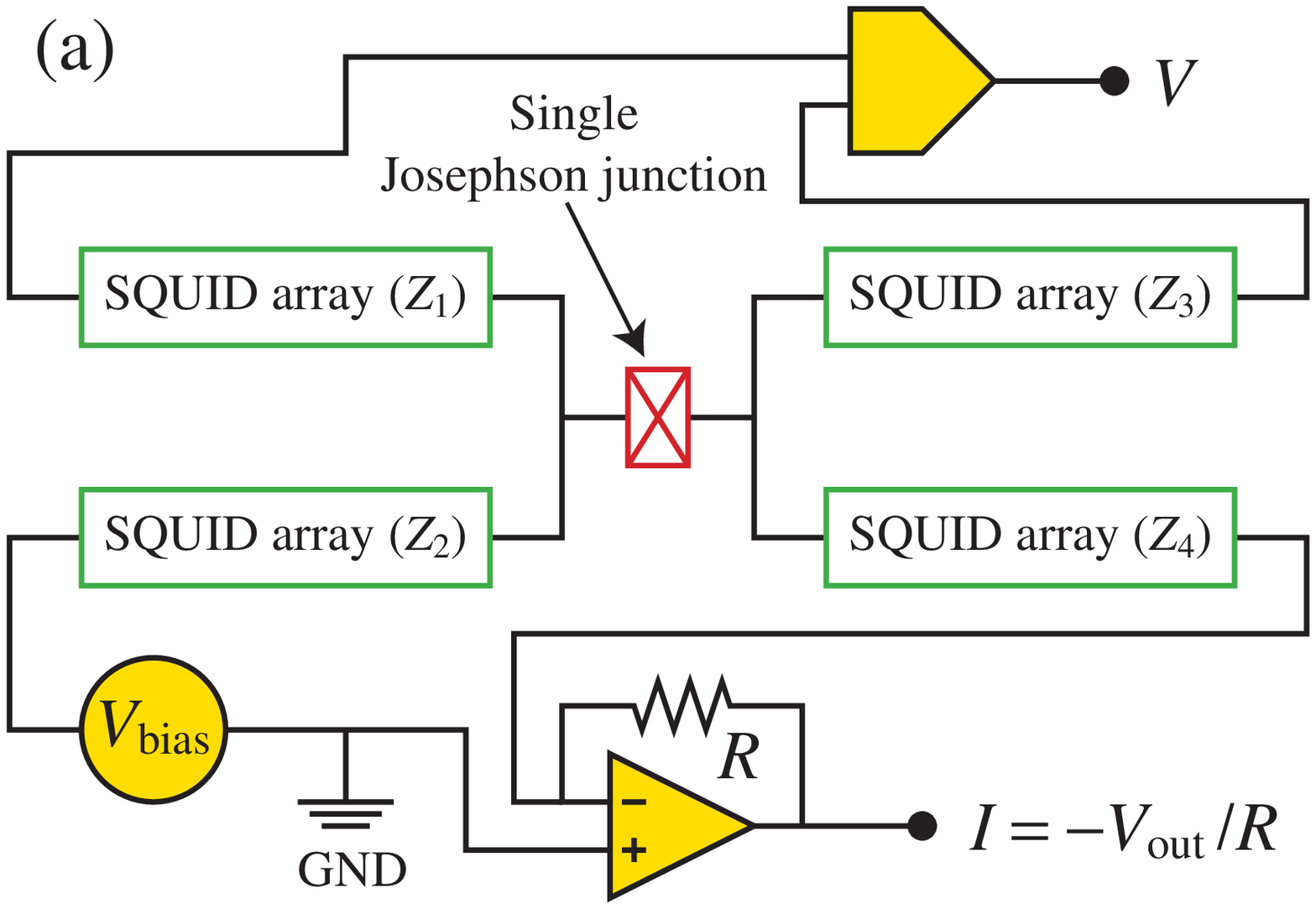}

\vspace{\baselineskip}

\includegraphics[width=0.9\columnwidth,clip]{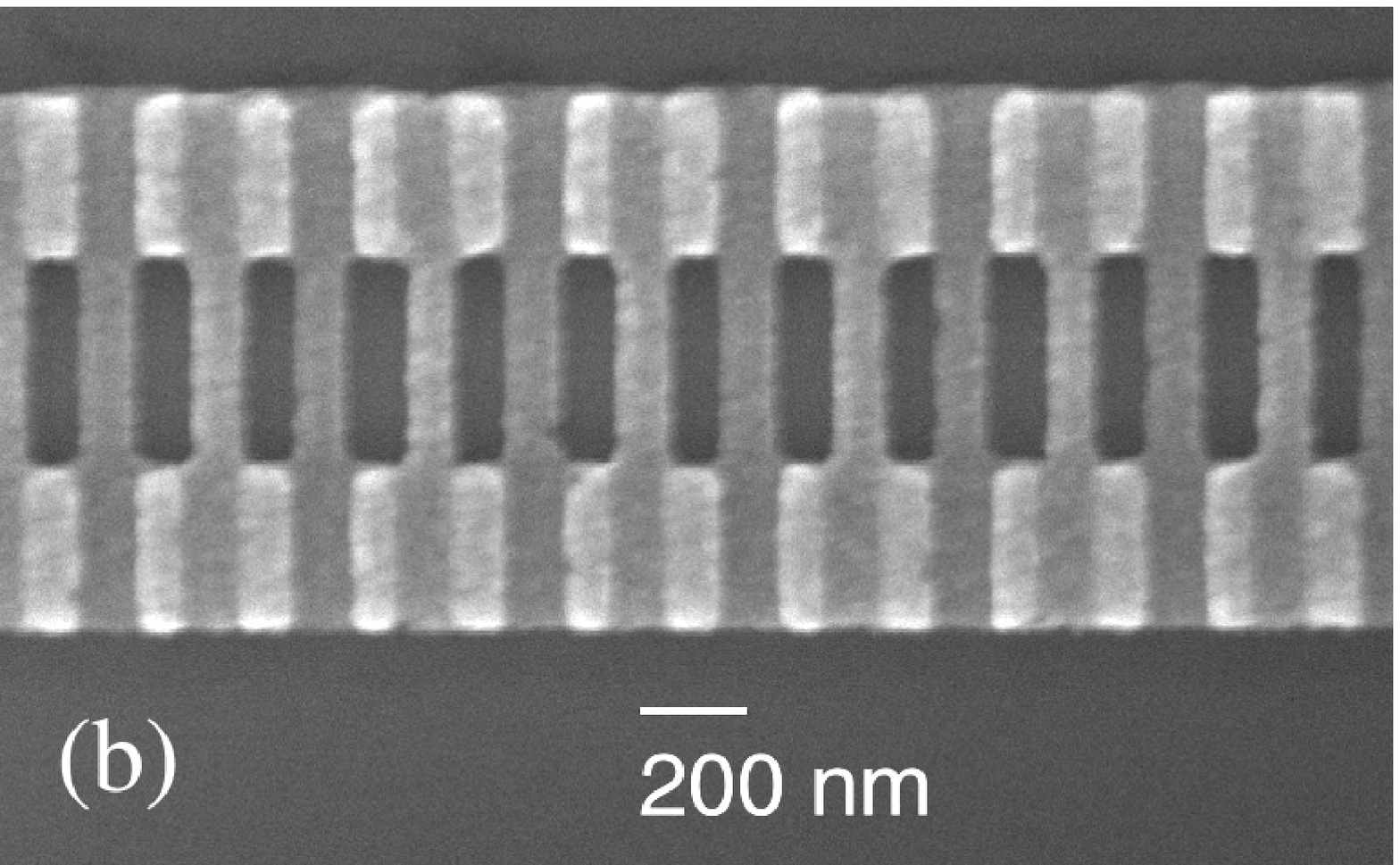}
%
%
\caption{\label{fig:sample}
(a) Schematic diagram of the samples and the circuit for measurements.  
(b) Scanning electron micrograph of a part of the 1D SQUID array.   
}
\end{figure}

A schematic diagram of the Al/Al$_2$O$_3$/Al 
samples is shown in Fig.~\ref{fig:sample}a.  
A single Josephson junction is at the center of Fig.~\ref{fig:sample}a.   
The junction area is 0.1$\times$0.1~$\mu$m$^2$.  
For a scanning electron micrograph of the single junction, see 
Fig.~1 of Ref.~\onlinecite{Wat01PRL}.  
On each side of the single junction, 
there are two leads enabling four-point measurements of the single junction.  
A part of each lead close to the single junction consists 
of a 1D array of dc SQUIDs.  
We show in Fig.~\ref{fig:sample}b a scanning electron micrograph 
of a part of the SQUID array.   
Each of the electrodes in the array is connected 
to its neighbors by two junctions in parallel, thus forming 
a dc SQUID between nearest neighbors.  
The area of each junction in the SQUID array is 0.3$\times$0.1~$\mu$m$^2$ 
and the effective area of the SQUID loop is 0.7$\times$0.2~$\mu$m$^2$.  
All of the samples have the same configuration, except for 
the number $N$ of junction pairs in each SQUID array.  
The samples are characterized by the normal-state resistance $R_n$ 
of the single junction, the normal-state resistance $r_n'$ 
per junction pair of the SQUID arrays, and $N$.  
The parameters of the samples are listed in Table~\ref{tab:sample}.  
\begin{table*}
\caption{
List of the samples.  
$R_n$ is the normal-state resistance  
of the single Josephson junction; 
$E_J$ is the Josephson energy; 
$E_C\equiv e^2/2C$ is the charging energy, 
where the capacitance $C$ is 
estimated from the junction area assuming 130~fF/$\mu$m$^2$; 
$V_b$ and $I_{\rm cr}$ are the blockade voltage and  
the crossover current, respectively (see Figs.~\ref{fig:calc}b 
and \ref{fig:32A1SJn}); 
$R_{\rm qp}$ is the quasiparticle resistance, which is 
estimated from the measured $I_{\rm cr}$ in Sec.~\ref{subsec:comp}; 
$N$ is the number of junction pairs in each SQUID-array lead; 
$r_n'$ is the normal-state resistance per junction 
pair of the SQUID-array leads.  
}
\label{tab:sample}
\begin{ruledtabular}
\begin{tabular}{cddddrrd}
 & \multicolumn{5}{c}{Single Josephson junction} & 
\multicolumn{2}{c}{Leads} \\ \cline{2-6} \cline{7-8}
Sample & \multicolumn{1}{c}{$R_n$ (k$\Omega$)} & 
\multicolumn{1}{c}{$E_J/E_C$} & 
\multicolumn{1}{c}{$V_b$ ($\mu$V)} & 
\multicolumn{1}{c}{$I_{\rm cr}$ (nA)} & 
\multicolumn{1}{c}{$R_{\rm qp}$ (k$\Omega$)} & 
\multicolumn{1}{c}{$N$} & 
\multicolumn{1}{c}{$r_n'$ (k$\Omega$)} \\
\hline
A & 1.0 & 9    & \multicolumn{1}{c}{$-$} & 
 \multicolumn{1}{c}{$-$} & \multicolumn{1}{c}{$-$} & 255 & 1.0 \\ 
B & 1.1 & 8    & \multicolumn{1}{c}{$-$} & 
 \multicolumn{1}{c}{$-$} & \multicolumn{1}{c}{$-$} & 255 & 1.1 \\  
C & 1.2 & 8    & \multicolumn{1}{c}{$-$} & 
 \multicolumn{1}{c}{$-$} & \multicolumn{1}{c}{$-$} & 255 & 1.2 \\  
D & 1.8 & 5    & \multicolumn{1}{c}{$-$} & 
 \multicolumn{1}{c}{$-$} & \multicolumn{1}{c}{$-$} & 65 & 0.6 \\   
E & 5.4 & 1.7  & \multicolumn{1}{c}{$-$} & 
 \multicolumn{1}{c}{$-$} & \multicolumn{1}{c}{$-$} & 33 & 0.5 \\   
F & 7.7 & 1.2  &  8 & 0.19 & $(0.6-1.4)\times10^3$ & 255 & 1.0 \\  
G & 9.0 & 1.0  & 10 & 0.19 & $(0.5-1.2)\times10^3$ & 255 & 1.1 \\ 
H & 12  & 0.8  & 11 & 0.08 & $(1-3)\times10^3$ &  33 & 1.2 \\ 
I & 17  & 0.5  & 25 & 0.08 & $(1-3)\times10^3$ &  65 & 1.4 \\ 
J & 31  & 0.3  & 29 & 0.02 & $(4-9)\times10^3$ &  65 & 1.9 \\ 
K & 41  & 0.21 & 28 & 0.02 & $(4-9)\times10^3$ & 255 & 5   \\ 
L & 58  & 0.15 & 29 & 0.01 & $(6-11)\times10^3$ & 255 & 7 \\
\end{tabular}
\end{ruledtabular}
\end{table*} 

It is difficult to determine the uniformity 
of these 1D SQUID arrays when only the series 
resistance of all junctions can be measured. 
Previous work\cite{Hav00} showed that the variations 
of the normal-state resistance per junction pair 
were less than 6\% for six arrays having nominally 
identical junctions made simultaneously on the same chip, 
with the number of junction pairs ranging from 7 to 255.  
Furthermore, critical-current switching at zero magnetic field 
for these arrays indicated a high degree of uniformity 
(Fig.~4 of Ref.~\onlinecite{Hav00}), 
and the magnetic-field dependence of the arrays was smooth.  
For the samples of this work, 
the uniformity of the SQUID arrays could be estimated 
by comparing the two arrays on the right- and left- hand sides 
of the single junction.  On an average, the normal-state 
resistance of the arrays agreed to within 5\%.  
However, if the tunnel-barrier thickness was  
identical for all junctions, one would expect 
$R_n/r_n'\approx6$ from the junction area 
for our samples.  In Table~\ref{tab:sample}, we see 
a large variation in  $R_n/r_n'(=1-17)$, which would be mainly 
due to the variation of $R_n$ for the following reasons:   
In order to obtain $r_n'$, we do not examine one junction pair 
in the arrays, but measure two arrays in series and divide 
by $2N$.  This kind of ``averaging" does not occur 
when we determine $R_n$.  Moreover, the single junction 
is much smaller than the junctions in the SQUID arrays, 
and small variations in the lithography and evaporation 
angle cause larger changes in the area of the smaller junction, 
thus leading to larger variations in the junction area 
and thereby the resistance and capacitance.

The tunnel junctions were fabricated on a SiO$_2$/Si substrate 
with electron-beam lithography and a double-angle 
shadow evaporation technique.\cite{Hav96}  
We employed a double-layer resist (ZEP520/PMGI) 
method in order to achieve a large enough undercut profile 
for the desired shift, which in our case was 0.2~$\mu$m.  
The electron-gun evaporation of Al was done at a rate of $0.1-0.4$~nm/s 
in a turbo-molecular-pumped vacuum system 
with the base pressure of $<2\times10^{-6}$~Pa.  
The vacuum system is equipped with an angle stage, 
and with the stage tilted to an angle $+\theta\approx 13^{\circ}$, 
Al was deposited to a thickness of 20~nm.  
After deposition of the base electrode, O$_2$ was introduced into 
the chamber to a pressure of $1.3-4.0$~Pa for a period of $2-10$~min.     
The oxidation conditions, or the thickness of the Al$_2$O$_3$ layer, 
determines $R_n$ and $r_n'$.  
After oxidation, the top layer of Al was deposited to a thickness 
of 30~nm with the stage tilted to an angle $-\theta$.  

In the above process, bonding pads ($0.3\times0.35$~mm$^2$) were
also fabricated simultaneously, i.e., the pads are 20+30=50~nm-thick Al.  
Onto the pads, 25~$\mu$m-diameter Al wires were wedge bonded.  

\subsection{Measurements}
The samples were measured in a $^{3}$He-$^{4}$He dilution refrigerator 
(Oxford Instruments, Kelvinox AST-Minisorb) at $T=0.02-0.6$~K.  
The temperature was determined by measuring the resistance 
of a ruthenium-oxide thermometer fixed at the mixing chamber.  
To measure the resistance of the thermometer, we used an ac resistance 
bridge (RV-Elekroniikka, AVS-47).  
Magnetic fields on the orders of $1-10$~mT were applied by means 
of a superconducting solenoid.  In this magnetic-field range, 
the temperature error of ruthenium-oxide thermometers due to the 
magnetoresistance is negligibly small (e.g., a typical value 
of the error at $T=0.05$~K is less than 0.1\%).\cite{Wat01Cryo}
The samples were placed inside a copper rf-tight box that was
thermally connected to the mixing chamber.  
All the leads entering the rf-tight box were low-pass filtered 
by 1~m of Thermocoax cable.\cite{Zor95}  

We measured the $I$-$V$ curve of the single junction 
in a four-point configuration (see Fig.~\ref{fig:sample}a).   
The bias was applied through one-pair of SQUID-array leads,  
and the potential difference was measured through the other pair 
of SQUID-array leads with a high-input-impedance 
instrumentation amplifier based on two operational amplifiers 
(Burr-Brown OPA111BM, $10^{13}-10^{14}$~$\Omega$ input impedance 
according to the data sheet) and a preamplifier [Stanford Research 
Systems (SRS) SR560].  The current was measured with a current 
preamplifier (SRS SR570).  
When the voltage drop at the SQUID arrays 
was much larger than that at the single junction, 
the single junction was practically current biased.  

The SQUID arrays could be measured in a two-point configuration 
(same current and voltage leads) on the same side of the single 
junction.  Note that the two arrays are connected in series 
and that current does not flow through the single junction.  
In order to obtain $R_n$ and $r_n'$, we measured at $T=2-4$~K 
(above the superconducting transition temperature of Al).

\section{Results and discussion}
\label{sec:RD}
\subsection{SQUID arrays: tunable electromagnetic environment 
for the single junction}
\begin{figure}
\includegraphics[width=0.95\columnwidth,clip]{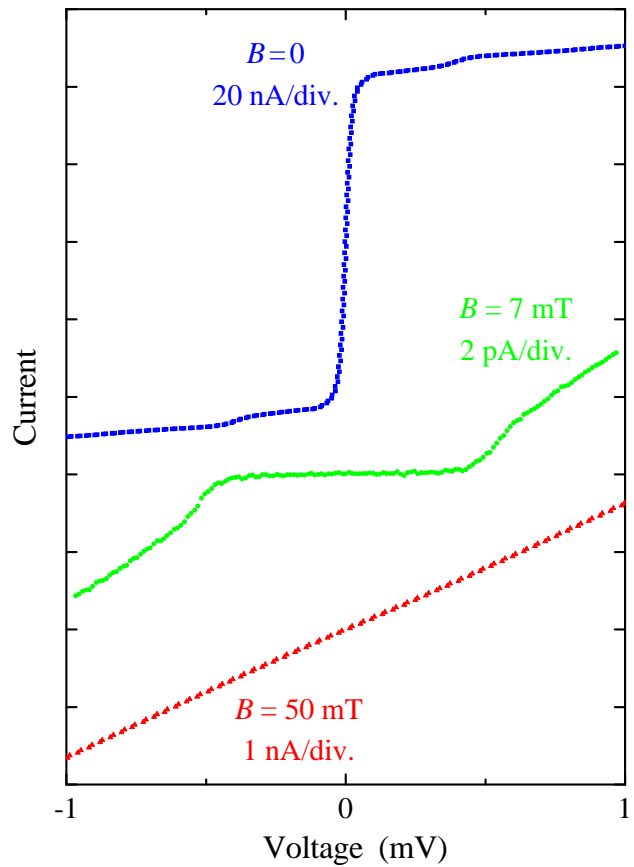}
\caption{\label{fig:IVleads}
Low-temperature ($T=0.05$~K) current-voltage characteristics 
of two SQUID-array leads connected in series 
at three magnetic fields for Sample~C. 
The origin of the current axis is offset 
for each curve for clarity.  
Note the difference in the current scale.  
}
\end{figure}

\begin{figure}
\includegraphics[width=0.95\columnwidth,clip]{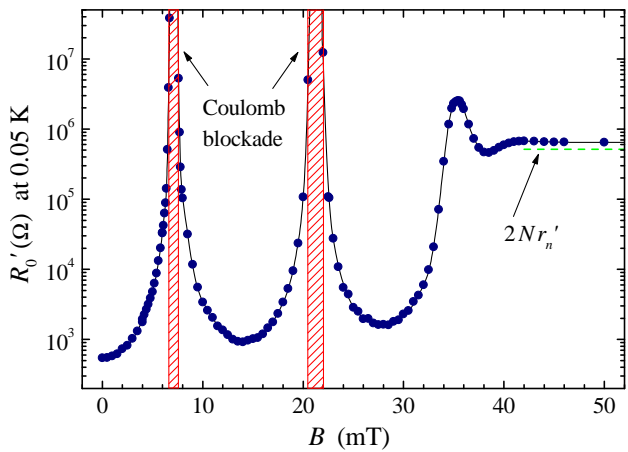}
\caption{\label{fig:R0leads}
Zero-bias resistance of the same SQUID-array leads 
as in Fig.~\ref{fig:IVleads} vs.\ external magnetic field 
at $T=0.05$~K. 
The solid curve is a guide to the eye.  
The shaded boxes and the broken horizontal line represent 
the region where Coulomb blockade was clearly seen 
in the current-voltage curve 
and the normal-state resistance of the two leads, $2Nr_n'$, 
respectively.   
}
\end{figure}

The effective Josephson energy $E_J'$ between adjacent islands 
in the SQUID arrays is modulated periodically by 
applying an external magnetic field $B$ perpendicular 
to the SQUID loop,  
\begin{equation}
\label{eq:EJ'}
E_J'\propto\left|\cos\left(\pi
\frac{\,BA\,}{\,\Phi_0\,}\right)\right|~,  
\end{equation}
so long as $B$ is sufficiently smaller than the critical field,    
where $A$ is the effective area of the SQUID loop 
and $\Phi_0= h/2e= 2\times10^{-15}$~Wb 
is the superconducting flux quantum.  
Figure~\ref{fig:IVleads} shows how the $I$-$V$ curve 
of two SQUID arrays on the same side 
of the single junction changes depending on $B$.    
The $I$-$V$ curve is Josephson-like at $B=0$, while 
a Coulomb blockade is developed at $B=7$~mT, 
where the normalized flux 
\begin{equation}
\label{eq:flux}
f\equiv\frac{\,BA\,}{\,\Phi_0\,} 
\end{equation}
is close to 1/2 for our samples with $A=0.14$~$\mu$m$^2$.  
Note that the current scale for $B=0$ is 
$10^4$ times larger than that for $B=7$~mT.  
In this manner, the $I$-$V$ curve is varied periodically  
until $B$ becomes comparable to the critical field.  
At $B=50$~mT, the superconductivity is suppressed considerably, 
and the $I$-$V$ curve is almost linear.  
The magnetic-field dependence of the SQUID arrays 
is summarized in Fig.~\ref{fig:R0leads}, 
where the zero-bias resistance $R_0'$ 
of the same SQUID arrays is plotted vs.\ $B$.  
As expected from Eq.~(\ref{eq:EJ'}), $R_0'$ oscillates 
in the low-field regime.  In the shaded regions, a Coulomb 
blockade is clearly observed and the maximum value of $R_0'$ 
can be as high as $10^9-10^{10}$~$\Omega$ (see Fig.~2 of 
Ref.~\onlinecite{Wat01PRL} for $R_0'$ of Sample~B 
at $0\leq B \leq 9$~mT).   
At $B>40$~mT, the oscillation is suppressed and $R_0'$ 
approaches the value in the normal state, $2Nr_n'$.  

In summary of this subsection, 
the SQUID arrays, which act as an electromagnetic 
environment for the single junction in this work, 
can be tuned over a wide range {\itshape in situ} 
with $f$.  
Henceforth, we use $f$ defined by Eq.~(\ref{eq:flux}) 
rather than $B$ for the purpose of indicating 
the external field.   

\subsection{Coulomb blockade induced in the single 
Josephson junction by tuning the environment}
\label{subsec:CB}
\begin{figure}
\includegraphics[width=0.95\columnwidth,clip]{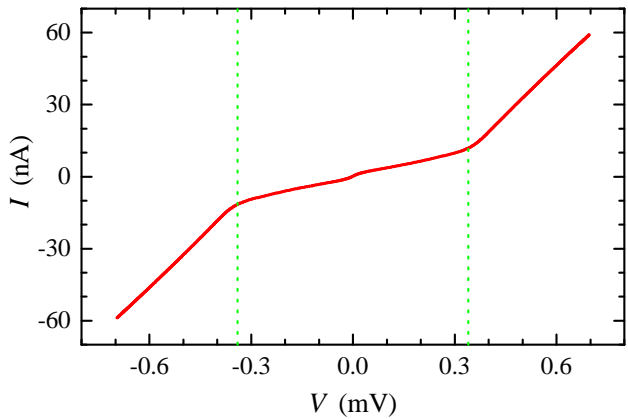}
\caption{\label{fig:32A1SJw}
Current-voltage characteristics 
of the single Josephson junction at $T=0.04$~K 
and $f=0$ for Sample~G.  
For the definition of $f$, see Eq.~(\ref{eq:flux})
The vertical broken lines represent 
the superconducting-gap voltage $\pm2\Delta_0/e$.   
}
\end{figure}

In this subsection, we take Sample~G for example.   
We look at the $I$-$V$ curve and the influence of 
the electromagnetic environment on it.  
(The $I$-$V$ curves of Sample~I are shown 
in Figs.~3--5 of Ref.~\onlinecite{Wat01PRL}.)  
Figure~\ref{fig:32A1SJw} shows the $I$-$V$ curves 
for the single junction at $f=0$.  
The vertical broken lines represent the superconducting-gap 
voltage $\pm2\Delta_0/e$, which is $\pm0.34$~mV 
for Al according to Ref.~\onlinecite{Kit96}.  
In a conventional Josephson junction ($E_J\gg E_C$), 
a finite current flows at $V=0$ 
as expected within the RCSJ (resistively and capacitively 
shunted junction) model.\cite{Tin96}  
This supercurrent has also been observed in 
small-capacitance Josephson junctions ($E_J\sim E_C$),  
and when a small-capacitance junction was biased with a 
carefully filtered low-impedance circuit,\cite{Ste01} 
the magnitude of the supercurrent reached the value 
predicted by the theory.
Our single Josephson junction  
is biased with SQUID arrays, whose $R_0'$ is much 
higher than the impedance ($\sim24$~$\Omega$) 
of the biasing circuit of Ref.~\onlinecite{Ste01}  
even at $f=0$.  
For the case of Fig.~\ref{fig:32A1SJw} (Sample~G at $f=0$), 
$R_0'=0.21$~k$\Omega$.   
Ingold and Grabert\cite{Ing99} considered how  
phase fluctuations suppress the supercurrent 
in a small-capacitance Josephson junction.  
The small capacitance of a junction is a source 
of phase fluctuations because the charge on the capacitance 
is the conjugate variable to the Josephson-phase difference.  
In addition, there is a strong relationship between 
the fluctuations and the electromagnetic environment.  
They assumed a purely Ohmic environment $R$, 
and showed that the supercurrent is rounded 
as $R$ is increased.  
In such a context, 
the shape of the $I$-$V$ curve in Fig.~\ref{fig:32A1SJw} 
may be explained for the region in the vicinity of 
$V=0$, although our SQUID arrays are not represented 
by a purely Ohmic impedance.   
At higher voltages, finite currents still 
flow in Fig.~\ref{fig:32A1SJw} even within the superconducting gap.  
Theoretically, at a finite voltage $|V|<2\Delta_0/e$, 
current can flow when the tunneling Cooper pairs can release   
their excess energy $2eV$, e.g., to the degrees of freedom 
in the electromagnetic environment.  
Quasiparticle excitations would also be responsible 
for the current especially at $|V|\sim2\Delta_0/e$.  
Below we will concentrate on the immediate vicinity of 
$V=0$ (typically $\pm30$~$\mu$V), where the $I$-$V$ curve is 
most sensitive to the state of the electromagnetic environment.    

\begin{figure}
\includegraphics[width=0.95\columnwidth,clip]{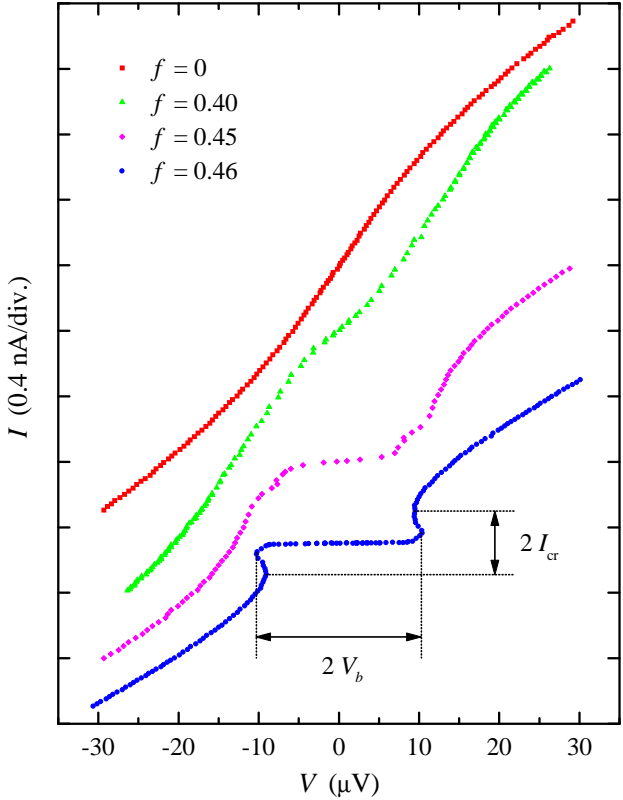}
\caption{\label{fig:32A1SJn}
Current-voltage ($I$-$V$) curves of the single Josephson junction  
in different environment at $T=0.04$~K for Sample~G.  
From top to bottom, the normalized flux defined 
by Eq.~(\ref{eq:flux}) is 0, 0.40, 0.45, 
and 0.46, respectively.   
The origin of the current axis is offset 
for each curve for clarity.  
}
\end{figure}

Figure~\ref{fig:32A1SJn} shows 
how the single-junction $I$-$V$ curve depends on 
the environment.  
As $f$ is varied, 
the curve develops a Coulomb blockade.  
As we have mentioned in Sec.~\ref{sec:intro}, 
the Josephson energy of the single junction is independent 
of $f$ because it does not have a SQUID configuration 
and the field $f\Phi_0/A<7$~mT applied here is much 
smaller than the critical field. 
The electromagnetic environment for the single junction 
(the SQUID arrays), however, is strongly varied with $f$.  
The behavior of the single junction demonstrated 
in Fig.~\ref{fig:32A1SJn} does not result from the magnetic-field 
influence on the single-junction $I$-$V$ curve, but rather from 
an environmental effect on the single junction.  This experiment 
demonstrates in a direct way that the single-junction   
$I$-$V$ curve is indeed sensitive to the electromagnetic environment.  

\begin{figure}
\includegraphics[width=0.95\columnwidth,clip]{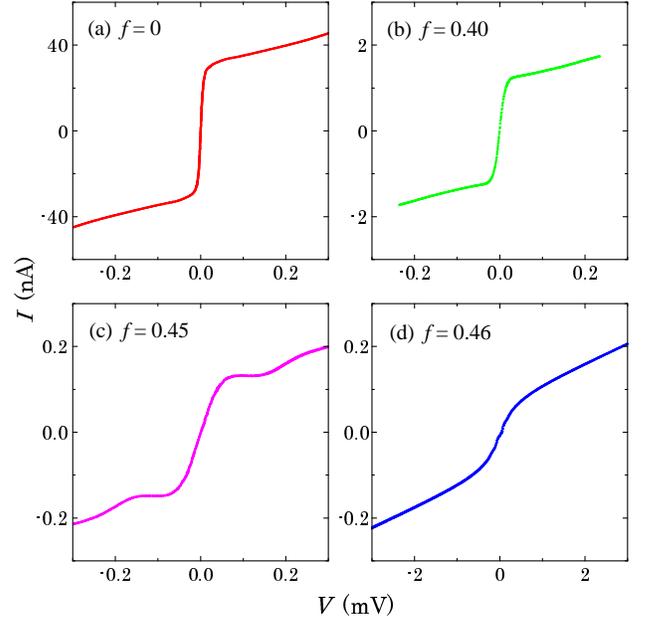}
\caption{\label{fig:32A1leads}
Current-voltage curves of the two SQUID-array 
leads connected in series at $T=0.04$~K for Sample~G.  
From (a) to (d), the normalized flux defined 
by Eq.~(\ref{eq:flux}) is 0, 0.40, 0.45, 
and 0.46, respectively.   
Note the difference in the current scale, 
and that the voltage scale for (d) is 
10 times larger than that for the others.  
}
\end{figure}

Figure~\ref{fig:32A1leads} shows the $I$-$V$ curves 
of the two SQUID-array leads connected in series 
at the same $f$ as in Fig.~\ref{fig:32A1SJn}.  
The $I$-$V$ curves of the leads are nonlinear, and in general the SQUID 
array is not described by a liner impedance model.\cite{Hav00}  
However, the environment may be characterized by $R_0'$.  
From (a) to (d) in Fig.~\ref{fig:32A1leads}, $R_0'=0.21$~k$\Omega$, 
11~k$\Omega$, 0.33~M$\Omega$, and 6.5~M$\Omega$, respectively.   
Coulomb blockade is distinct only when $R_0'\gg R_K$, 
which is consistent with the theoretical 
conditions for the clear observation of Coulomb blockade 
in single junctions.\cite{Ing92}  
For an arbitrary linear environment characterized by $Z(\omega)$, 
$\mbox{Re}[Z(\omega)]\gg R_K$ is required for the Coulomb blockade 
of single-electron tunneling and $\mbox{Re}[Z(\omega)]\gg R_Q$ 
for that of Cooper-pair tunneling.\cite{Ing92}  
It is interesting to note that while the $I$-$V$ curve 
of the leads is ``Josephson-like" (differential resistance is 
lower around $V=0$) at all $f$ shown in 
Fig.~\ref{fig:32A1leads}, that of the single junction 
is ``Coulomb-blockade-like" at $f=0.40$ and   
exhibits a clear Coulomb blockade at $f\geq0.45$.

\subsection{Comparison with the numerical calculation}
\label{subsec:comp}
\begin{figure}
\includegraphics[width=0.95\columnwidth,clip]{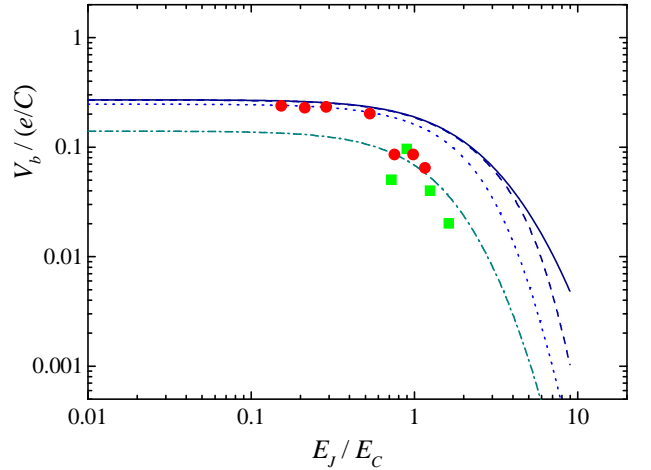}
\caption{\label{fig:Vb}
Blockade voltage $V_b$ divided by $e/C$ as a function 
of $E_J/E_C$. 
From top to bottom, the curves represent 
the numerical calculations for normalized temperatures  
$k_BT/E_C=0$, 0.02, 0.1, and 0.5, respectively.  
The boxes represent the samples with the nominal junction 
area of 0.01~$\mu$m$^2$ in Ref.~\onlinecite{Hav91}.  
}
\end{figure}

The region of negative differential resistance seen 
at the bottom curve of Fig.~\ref{fig:32A1SJn}  
is related to coherent tunneling of single Cooper pairs 
according to the theory\cite{Ave91,Sch90} 
of a current-biased single Josephson junction in an environment 
with sufficiently high impedance.  
Following Ref.~\onlinecite{Gei88}, 
we have calculated the blockade voltage $V_b$ 
numerically as a function of $E_J/E_C$.\cite{Wat01SUST}    
The numerical results are compared with the 
measured $V_b$ in Fig.~\ref{fig:Vb}.
The data from Ref.~\onlinecite{Hav91} for the samples 
with the same nominal junction area ($0.1\times0.1$~$\mu$m$^2$) 
as in this work are also plotted.  
For Sample~G we used the data at $f=0.46$ (the bottom curve 
in Fig.~\ref{fig:32A1SJn}) in order to obtain $V_b$.  
At $f=0.46$ the voltage drop at the SQUID arrays is $10^2$ times 
larger than that at the single junction, and the single junction 
is therefore considered to be current biased.  Compare the 
voltage scale of Figs.~\ref{fig:32A1SJn} and \ref{fig:32A1leads}d.  
We calculated $E_J$ from $R_n$, 
\begin{equation}
E_J=\frac{\,h\Delta_0\,}{\,8e^2R_n\,}~.  
\end{equation}
For $E_C$ we employed $c_s=130$~fF/$\mu$m$^2$, and with this value 
the experimental data, especially those for this work 
(the circles in Fig.~\ref{fig:Vb}), 
agree with the numerical calculation.  
Actually, a smaller value, 
$c_s=45\pm5$~fF/$\mu$m$^2$~(Ref.~\onlinecite{Lic89}), 
which was obtained for the junctions with $3\times28$~$\mu$m$^2$ 
and $7\times54$~$\mu$m$^2$, has been frequently 
employed.\cite{Cho98,Hav00,Hav91,Shi97,Hav01}  
Our apparently large $c_s$ may be partly explained 
by distributed capacitance of the SQUID arrays  
or by the non-infinite impedance of the environment 
(SQUID arrays), which would have the effect of 
shunting the single junction with an additional 
capacitance (see Fig.~\ref{fig:circuits}).  
We also note that the uncertainty in $c_s$ 
seems to be large when the junction area 
is on the order of 0.01~$\mu$m$^2$ or smaller, 
as we have discussed in Ref.~\onlinecite{Wat01PRL}.  

\begin{figure}
\includegraphics[width=0.95\columnwidth,clip]{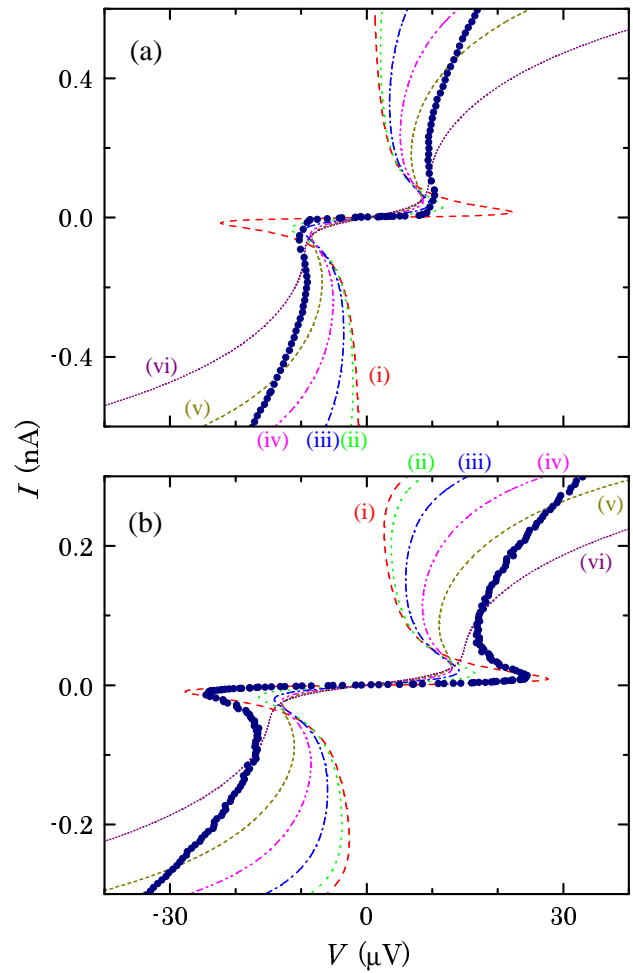}
\caption{\label{fig:IVcomp}
Current-voltage curves of single Josephson junctions.  
(a) The measured curve (solid circles) of Sample~G 
at $T=0.04$~K and $f=0.46$ is compared 
with the numerical calculations for $E_J/E_C=0.53$ 
and $\alpha=0.006$ [see Eqs.~(\ref{eq:alpha}) 
and (\ref{eq:flux}) for the definition of $\alpha$ 
and $f$, respectively].  
(b) Sample~I at $T=0.02$~K and $f=0.49$ with 
the calculatins for $E_J/E_C=0.99$ and $\alpha=0.003$.  
For both the figures, $C=1.3$~fF is assumed 
in the calculations, and from (i) to (vi), 
$k_BT/E_C=0.05$, 0.3, 0.4, 0.5, 0.6, and 0.8, 
respectively.  
}
\end{figure}

Fixing $C=1.3$~fF, we tried to reproduce the 
measured $I$-$V$ curve by optimizing the other 
parameters of the numerical calculation, $\alpha$ 
and $k_BT/E_C$.  
Examples are shown in Fig.~\ref{fig:IVcomp}, 
where we adjusted $\alpha$ roughly 
by considering the current scale.   
For both Samples~G and I, the slope of the experimental data 
at $|V|<V_b$ and $I\sim0$ can be fit by adjusting $\alpha$, 
so that it is consistent with the theoretical curve for 
$k_BT/E_C=0.05$ ($T=0.036$~K for $C=1.3$~fF), which is close 
to the temperature of the mixing chamber, $0.02-0.04$~K.    
At higher bias currents, however, 
the theory predicts for $k_BT/E_C=0.05$ a ``back bending" 
much bigger than that in the experimental data.     
In terms of the magnitude of ``back bending," 
the curves for higher temperatures, 
$k_BT/E_C=0.3-0.6$ ($T=0.21-0.43$~K), 
are more similar to the data.  ``Back bending" in the 
theoretical curve disappears when the temperature is raised 
to $k_BT/E_C=0.8$ ($T=0.57$~K).  
Although the theory explains the general shape of the measured 
$I$-$V$ curve, it does not yield a complete agreement.  
It would be reasonable to assume that the effective 
temperature of the single junction depends on the bias current, 
and increases with increasing current due to heating.   
This assumption reduces the discrepancy between the experimental 
data and the theory.  
Another possible origin of the discrepancy would be 
the imperfect isolation of the single junction.  
While the theory assumes that the single junction is 
completely isolated from the environment, we have 
biased the single junction with the SQUID arrays 
having non-infinite $R_0'$.  
Moreover, the SQUID arrays are a nonlinear environment.  
  
\begin{figure}
\includegraphics[width=0.95\columnwidth,clip]{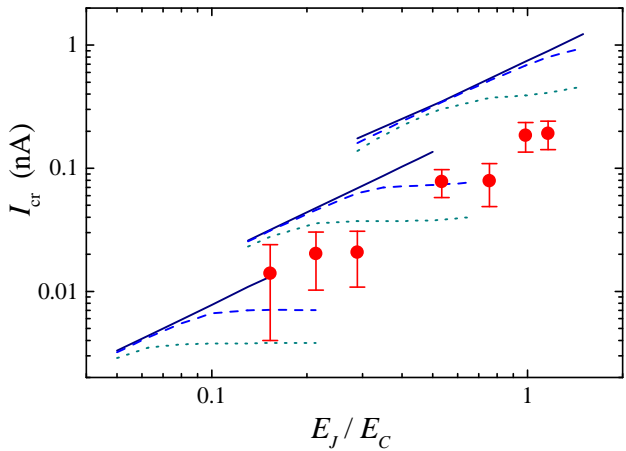}
\caption{\label{fig:Icr}
Crossover current $I_{\rm cr}$ vs. $E_J/E_C$.  
The curves represent 
the numerical calculations for $C=1.3$~fF, and from top to bottom  
$(\alpha,~k_BT/E_C)=(10^{-2},0)$, 
$(10^{-2},0.3)$, $(10^{-2},0.5)$, $(10^{-3},0)$, 
$(10^{-3},0.3)$, $(10^{-3},0.5)$, $(10^{-4},0)$, 
$(10^{-4},0.3)$, and $(10^{-4},0.5)$, respectively.
See Eq.~(\ref{eq:alpha}) for the definition of $\alpha$.  
}
\end{figure}

Finally in this subsection, we estimate $R_{\rm qp}$ 
from $I_{\rm cr}$ based on the theory.  
We plot the measured 
$I_{\rm cr}$ as a function of $E_J/E_C$ in Fig.~\ref{fig:Icr}
together with some theoretical curves 
based on our numerical calculation.  
For the estimate of $R_{\rm qp}$ we used the numerical 
results for $k_BT/E_C=0.3$ and 0.5 ($T=0.21$ and 0.36~K),  
because with these values the numerical calculation gave  
an $I$-$V$ curve similar to the experimental one 
around $I_{\rm cr}$ as we have seen in Fig.~\ref{fig:IVcomp}.     
The estimated $R_{\rm qp}$ is listed in Table~\ref{tab:sample}.  
The order of $R_{\rm qp}$ ($R_{\rm qp}=10^0-10^1$~M$\Omega$) 
and the ratio to $R_n$ ($R_{\rm qp}/R_n=10^2-10^3$) 
are within a reasonable range.  

\subsection{Superconductor-insulator transition in single 
Josephson junctions driven by the environment}
\begin{figure}
\includegraphics[width=0.95\columnwidth,clip]{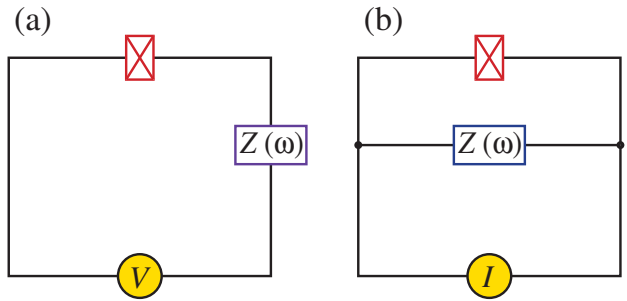}
\caption{\label{fig:circuits}
(a) Small-capacitance single Josephson junction 
biased by an ideal voltage source with an electromagnetic 
environment $Z(\omega)$ in series.  
(b) An equivalent circuit to (a) when $I=V/Z(\omega)$, 
where the single junction is biased 
by an ideal current source and 
shunted by $Z(\omega)$.  
}
\end{figure}

The numerical calculation in the preceding subsection 
agrees with the experiment qualitatively, 
however, the apparently large $c_s$ and the discrepancies in 
Fig.~\ref{fig:IVcomp} suggest 
that the single junction would still be dependent 
on the environment even at $R_0'\gg R_K$.   
The influence of a linear electromagnetic environment $Z(\omega)$ 
on small-capacitance Josephson junction 
has been studied in the framework of the perturbation 
theory\cite{Fal91,Ing92,Ing99} by modeling the whole circuit 
as in Fig.~\ref{fig:circuits}a.  
By assuming an Ohmic environment $Z(\omega)=R$, 
the Cooper-pair current-voltage characteristics have been 
calculated for $E_J/E_C\ll1$.  
The theory predicts Coulomb blockade of Cooper-pair tunneling 
for $R_Q/R\ll1$, and supercurrent at or in the vicinity 
$V=0$ for $R_Q/R\gg1$.  
By comparing Fig.~\ref{fig:sample}a with Fig.~\ref{fig:circuits}a, 
we realize that the total impedance $Z_t$ seen by the junction 
in Fig.~\ref{fig:sample}a corresponds to $Z$ 
in Fig.~\ref{fig:circuits}a.  
If we assume in Fig.~\ref{fig:sample}a that 
the SQUID arrays alone determine the effective environment 
for the single Josephson junction at relevant frequencies, 
where all the four leads are shunted at the far ends of 
the SQUID arrays, and that the SQUID array is described 
by a linear impedance, 
we obtain $Z_t=
\left(Z_1^{-1}+Z_2^{-1}\right)^{-1}
+\left(Z_3^{-1}+Z_4^{-1}\right)^{-1}\approx Z_1$ 
(all the SQUID arrays are nominally the same).   
Note that the dc $I$-$V$ curve is the result of high-frequency quasicharge 
dynamics of the single junction, and thus one must consider what 
the junction sees at high frequencies.  The large capacitance 
introduced by the cryostat leads will effectively shunt the sample 
with a low impedance at the relevant frequencies. 
This is why we fabricated the SQUID arrays in the immediate 
vicinity of the single junction in order to avoid the lead 
capacitance shunting the single junction, and this is why 
we assume that the SQUID arrays alone 
determine the environment for the single junction.
We have been characterizing the environment by the zero-bias 
resistance of two arrays measured in series, $R_0'$.  
Since $Z_t\approx Z_1$, it would be more appropriate 
to use $R_0'/2$.  The factor of 2, however, 
is not very important in this work, and thus we keep 
using $R_0'$.  
We should note here that $Z\sim R_0'$ is a bold approximation.  
The $I$-$V$ curves of the SQUID arrays 
are nonlinear as we have seen in Figs.~\ref{fig:IVleads} and 
\ref{fig:32A1leads}, and in general the SQUID array 
is not described by a liner impedance.  
The frequency dependence\cite{Hav00} is also neglected 
in the approximation.  

\begin{figure}
\includegraphics[width=0.95\columnwidth,clip]{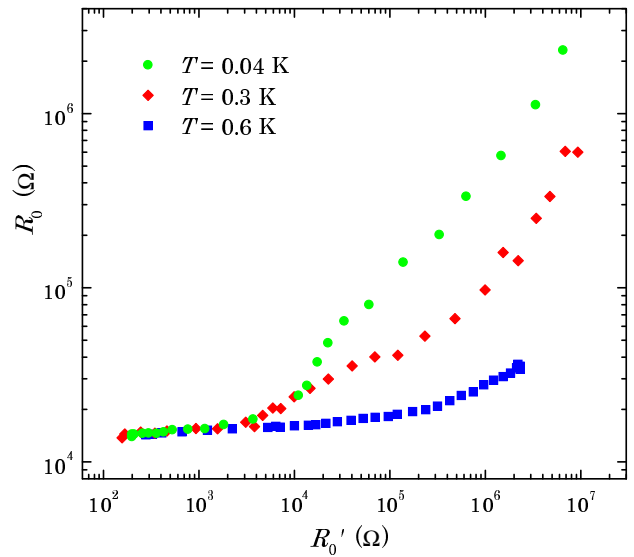}
\caption{\label{fig:TdepR0}
Zero-bias resistance $R_0$ of the single Josephson junction 
vs.\ zero-bias resistance $R_0'$ of the SQUID-array leads 
at $T=0.04$, 0.3, and 0.6~K for Sample~G.  
}
\end{figure}

In order to examine the $R_0'$ dependence further, we plot 
in Fig.~\ref{fig:TdepR0} the zero-bias resistance $R_0$ 
of the single junction vs.\ $R_0'$ 
at three different temperatures for Sample~G.  
At all the temperatures $R_0$ is a smooth function of $R_0'$  
and increases with increasing $R_0'$, 
which supports the notion that the approximation $Z\sim R_0'$ 
is adequate. 
Even at the largest $R_0'$, $R_0$ is still increasing 
with no sign of saturation.  
The top left point in Fig.~\ref{fig:TdepR0} for $T=0.04$~K 
and $R_0'=6.5$~M$\Omega$ ($f=0.46$) corresponds 
to the bottom $I$-$V$ curve in Fig.~\ref{fig:32A1SJn}.  
Thus, $R_0$ vs.\ $R_0'$ also suggests 
that the single junction would still be influenced 
by the environment 
at $R_0'\gg R_K$.       
The change of $R_0$ is weak at $R_0'<10^3-10^4$~$\Omega$.  
This fact would be understood in the following context.  
In the limit of $R_0'\rightarrow0$, the impedance of the 
electromagnetic environment for the Josephson junction at 
relevant frequencies is determined by the microwave 
impedance of the leads, which is usually of the order 
of the free-space impedance $Z_0\approx377$~$\Omega$.  
Thus, when $R_0'$ is of the order of $Z_0$ or smaller, 
our assumption that the SQUID arrays determine the environment 
is not valid, and $Z\sim Z_0$ irrespective of the value of $R_0'$.  
In Ref.~\onlinecite{Ste01}, special care was taken 
in order to achieve $Z/Z_0<0.1$ at all relevant frequencies.    

The interaction with a dissipative environment, 
or the dissipative dynamics of single Josephson junctions 
has also been studied theoretically\cite{Sch90,Kat00,Chu02} 
in the circuit shown in Fig.~\ref{fig:circuits}b with $Z(\omega)=R$.  
Note that the two circuits in Fig.~\ref{fig:circuits} 
are equivalent when $I=V/Z(\omega)$.  
As the dissipation is increased, i.e., $R$ is decreased, 
the Josephson junction undergoes a transition from an insulator to 
a superconductor.  The phase diagram for this SI transition 
has been derived\cite{Sch90} and qualitatively supported 
by the experiments.\cite{Yag97,Pen01}   
The phase (``superconductor" or ``insulator") is usually determined by 
the $I$-$V$ curve (or $dV/dI$ vs.\ $I$ curve\cite{Pen01}) 
at the lowest temperatures and/or the temperature dependence 
of the zero-bias resistance.\cite{Yag97}
We also find evidence of the SI transition 
in the nonlinearity of the $I$-$V$ curve.  For Sample~G, 
shown in Fig.~~\ref{fig:32A1SJn}, 
we see that the $I$-$V$ curve is  ``Josephson-like" 
when the leads have $R_0'=0.21$~k$\Omega$ 
($f=0$), whereas at $R_0'\geq0.33$~M$\Omega$ ($f\geq0.45$), 
there is a distinct Coulomb blockade.  
The temperature dependence of $R_0$ 
also supports the existence of the SI transition.  
We see in Fig.~\ref{fig:TdepR0} that 
at $R_0'<10^3-10^4$~$\Omega$, $R_0$ has little temperature 
dependence, which is similar to the behavior of a 1D 
system\cite{Cho98,Hav00,Hav01} in the superconducting phase.  
At $R_0'>10^4-10^5$~$\Omega$, on the other hand, 
we see insulating behavior, where $R_0$ increases rapidly as 
the temperature is lowered.  

\begin{figure}
\includegraphics[width=0.95\columnwidth,clip]{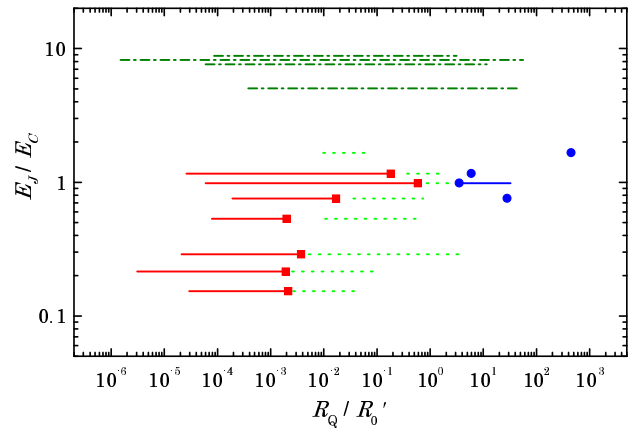}
\caption{\label{fig:phase}
Phase diagram for the superconductor-insulator 
transition in single Josephson junctions 
biased with tunable SQUID arrays.  
The transition is driven by the electromagnetic 
environment characterized by the zero-bias 
resistance $R_0'$ of the SQUID arrays.  
The horizontal axis is the ratio of the quantum 
resistance $R_Q\equiv h/(2e)^2\approx6.5$~k$\Omega$ 
to $R_0'$.  
The vertical axis is the ratio of the Josephson 
energy $E_J$ to the charging energy $E_C$ 
in the single junction.  
See text for the 
meaning of the lines and symbols.  
}
\end{figure}

\begin{figure}
\includegraphics[width=0.95\columnwidth,clip]{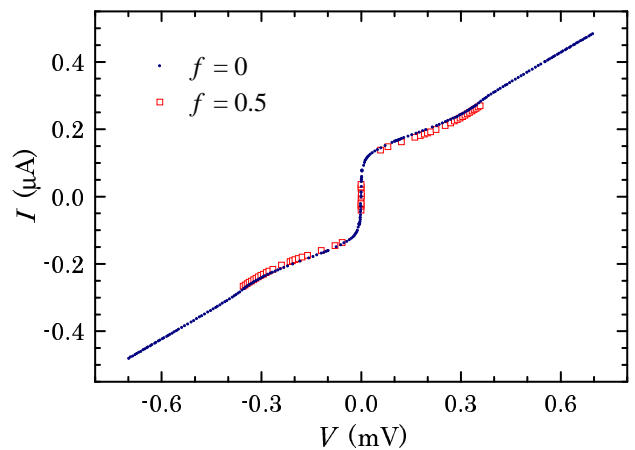}
\caption{\label{fig:26A2SJ}
Current-voltage characteristics 
of the single Josephson junction 
in two different environment ($f=0$ and 0.5) 
at $T=0.04$~K for Sample~D.  
For the definition of $f$, see Eq.~(\ref{eq:flux}).  
}
\end{figure}

We show in Fig.~\ref{fig:phase} the phase diagram for the 
SI transition determined by the $I$-$V$ curves at the lowest 
temperatures of all the samples studied in this work.  
Note that $R_0'$ was varied over as large as 
nine orders of magnitude in this work.  
In Samples~F--H, we observed the SI transition.  
The squares and the solid lines to the left of the 
squares indicate that insulating behavior 
(Coulomb blockade or ``Coulomb-blockade-like" $I$-$V$ 
curve) was observed.  
The circles and the solid lines to the right of the 
circles represent the superconducting 
phase (``Josephson-like" $I$-$V$ curve).  
In the range denoted by broken lines, the $I$-$V$ curve 
was almost linear, and it was hard to determine the phase.   
For Samples~I--L (Sample~E), the insulating 
(superconducting) phase alone was identified.  
Samples~A--D have a large $E_J/E_C$ ($\geq5$), and we expect 
from Fig.~\ref{fig:Vb} that their $V_b$ is too small 
($\ll1$~$\mu$V) to be detected in our dc measurements, 
i.e., the insulating phase is never identified.    
Indeed, the $I$-$V$ curve of Samples~A--D is almost 
independent of $R_0'$.  As an example, we show 
in Fig.~\ref{fig:26A2SJ} the $I$-$V$ curve of 
Sample~D at $R_0'=0.14$~k$\Omega$ ($f=0$) and 
$R_0'=2$~M$\Omega$ ($f=0.5$).  
It is Josephson-like and in a contrast 
with the curve in Fig.~\ref{fig:32A1SJw}.  
We do not discuss the SI transition for Samples~A--D, 
and just indicate the $R_0'$ range covered in the 
measurements by dot-dashed lines in Fig.~\ref{fig:phase}.  
According to the theory\cite{Sch90} of single Josephson junctions 
with an Ohmic shunt $R$, the phase boundary is located 
at $R_Q/R=1$ irrespective of the ratio $E_J/E_C$ 
when the quasiparticle tunneling is not important ($\alpha\ll1$) 
and the transition is driven solely by $R$.     
In Samples~F--L, $\alpha=10^{-4}-10^{-2}\ll1$ 
from Table~\ref{tab:sample}.  
It is interesting to see that Fig.~\ref{fig:phase}, 
the phase diagram for our system with {\itshape non-Ohmic} 
environment, is consistent with the theoretical phase 
diagram where an {\itshape Ohmic} shunt is assumed.    

\section{Conclusion}
We have studied the current-voltage ($I$-$V$) characteristics 
of single Josephson junctions biased with a tunable 
electromagnetic environment composed of SQUID arrays.  
We have demonstrated that the single junction 
is indeed sensitive to the state of the environment, 
and a Coulomb blockade is induced in the single junction 
by increasing the zero-bias resistance $R_0'$ 
of the SQUID arrays.  
When $R_0'$ is much higher than the quantum resistance 
$R_K\equiv h/e^2\approx26$~k$\Omega$,   
a region of negative differential resistance has also been 
observed in the $I$-$V$ curve of the single junction.   
The negative differential resistance is evidence of coherent 
single-Cooper-pair tunneling according to the theory 
of current-biased single Josephson junctions.  
Based on the theory we have calculated the $I$-$V$ curves 
numerically.  The calculation has reproduced the measured 
$I$-$V$ curves at $R_0'\gg R_K$ qualitatively.  
We have also discussed the superconductor-insulator (SI) 
transition in single Josephson junction driven by the 
environment.  We have characterized the environment 
by $R_0'$ and determined the phase diagram 
for the transition.  This phase diagram is consistent 
with the theoretical one for the SI transition 
in single Josephson junctions driven by an 
Ohmic shunt alone.  

\begin{acknowledgments}
We are grateful to R. L. Kautz for great help in the numerical 
calculation, and to T. Kato and F. W. J. Hekking for fruitful 
discussions.  This work was supported by Swedish NFR, 
and Special Postdoctoral Researchers Program and President's Special 
Research Grant of RIKEN.  
M. W. would like to thank the Japan Society for the Promotion 
of Science (JSPS) and the Swedish Institute (SI) 
for financial support.  
\end{acknowledgments}

\end{document}